\documentclass[aps,twocolumn,prb,amsmath,amssymb]{revtex4-1}
\usepackage{graphicx}
\usepackage{dcolumn}
\usepackage{color}
\usepackage{array}
\usepackage{multirow}
\usepackage[latin1]{inputenc}
\usepackage{graphics}
\usepackage{overpic}
\usepackage{bm}
\usepackage{subfigure}

\usepackage[english]{babel}
\begin{document}

\title{Properties of the ferroelectric visible light absorbing semiconductors:
Sn$_2$P$_2$S$_6$ and Sn$_2$P$_2$Se$_6$}

\author{Yuwei Li}
\affiliation{Department of Physics and Astronomy, University of Missouri, Columbia, MO 65211-7010 USA}
\author{David J. Singh}
\email{singhdj@missouri.edu}
\affiliation{Department of Physics and Astronomy, University of Missouri, Columbia, MO 65211-7010 USA}

\date{\today}

\begin{abstract}
Ferroelectrics with suitable band gaps have recently attracted attention
as candidate solar absorbing materials for photovoltaics.
The inversion symmetry breaking may promote
the separation of photo-excited carriers and allow voltages higher than the
band gap. However, these effects are not fully understood, in part
because of a lack of suitable model systems for studying these effects
in detail.
Here, we report properties of ferroelectric
Sn$_2$P$_2$S$_6$ and Sn$_2$P$_2$Se$_6$ using first principles calculations.
Results are given for the electronic structure, carrier pocket shapes,
optical absorption and transport.
We find indirect band gaps of
2.20 eV and 1.55 eV, respectively,
and favorable band structures for carrier transport,
including both holes and electrons.
Strong absorption is found above the direct gaps of 2.43 eV and 1.76 eV.
Thus these compounds may serve as useful model systems for understanding
photovoltaic effects in ferroelectric semiconductors.

\end{abstract}

\maketitle

\section{Introduction}

In conventional photovoltaic (PV) devices, electron$-$hole pairs are created
by light absorption and separated by the electric field in a depletion region.
The voltage is limited by the band gap.
Generally, large band gap photovoltaic devices,
which can have relatively high voltage,
have lower efficiency because of lower
visible light absorption.
Ferroelectric photovoltaic devices\cite{ferroelectric}
have been reported since 1962.
The photovoltaic charge separation mechanism is apparently different
from  conventional photovoltaics and the voltage can be significantly
higher than the band gap\cite{PV1ferro,grinberg2013perovskite}.
For example, in BiFeO$_3$\cite{seidel2011efficient}
the origin of photovoltaic effect was discussed
in terms of electron-hole separation at ferroelectric domain walls.
Thus it is claimed that inversion symmetry breaking in ferroelectric materials
associated with the spontaneous polarization produces an
asymmetry that promotes
the separation of photo-excited carriers and allows
voltages higher than the band gap.
The implication
of these higher voltages, if not compensated by other
losses, is that useful devices with properties distinct from conventional
p-n junction solar cells may be possible.
However, questions remain, for example,
whether the voltage can be maintained in a long duration steady state.

We also note that independent of the
charge separation mechanism, high defect tolerance is
beneficial in solar absorbers.
High dielectric constant is one mechanism for obtaining high defect
tolerance, favorable for carrier collection,
similar to high dielectric constant radiation detection
and energy materials, where the high dielectric constant screens
traps and reduces carrier scattering.
\cite{du2010enhanced,PhysRevApplied.7.024015,shuai}

Clearly, photovoltaic effects in ferroelectric materials have attracted
considerable recent attention both to understand these effects and because
of potential applications
\cite{PV1ferro,PV2ferro,cao2012high,qin2008high,glass1974high}. 
However, this has been impeded by
the lack of good well characterized ordered ferroelectric
semiconductor materials having band gaps in the visible or near infrared
that could be used in studies to understand ferroelectric PV effects and
the performance that can be obtained from them.

Here we report detailed investigation of the electronic,
optical and related properties of two Sn chalcogenide semiconductor
ferroelectrics Sn$_2$P$_2$S$_6$ and Sn$_2$P$_2$Se$_6$ with band gaps in
in the visible or near infrared range that may be useful in this context.
Prior work on Sn$_2$P$_2$S$_6$ has focused mainly on ferroelectric
properties, photorefraction, possible memory applications,
and characterization of the absorption edges in relation
to ferroelectricity
\cite{
Es_Sn2P2S6,
ah_shchennikov2011colossal,
aj_grigas2009xps,
x_carpentier,
am_tyagur2006spontaneous,
j_gamernyk,p_potuuvcek,
b_studenyak,
h_slivka,
j_gamernyk,
o_ruediger,
ae_kojima,
ao_potuuvcek2004luminescence,
ap_shumelyuk2010light,
ab_arnautova,
S6opt2,
S6opt3,
S6opt4,
S6opt5,
aa_shumelyuk,
ae_kojima}.
Sn$_2$P$_2$Se$_6$ has been less studied, but again band edges and
ferroelectric properties have been characterized.
\cite{
e_Sn2P2Se6,
h_slivka,
q_lipavivcius,
aq_dmitruk2007luminescence}.
The ferroelectric ordering temperatures are $T$c = 337 K,
and 193 K, for the sulfide and selenide, respectively.

Sn$_2$P$_2$S$_6$ and Sn$_2$P$_2$Se$_6$
are known ferroelectric materials.
Also, they are stoichiometric compounds, as distinct from alloys or
superlattice structures.
This removes the difficulty of growing and studying superlattices
or complications from alloys, and therefore may facilitate detailed
investigation of ferroelectric PV phenomena,
as well as comparison of experimental theoretical results for the
materials properties of importance for PV application.
They also have sizable polarization. The polarization of Sn$_2$P$_2$Se$_6$
from hysteresis loop measurements is $\sim$35 $\mu$C/cm$^2$ at low temperature,
\cite{b_studenyak}
which is comparable to BaTiO$_3$ (34 $\mu$C/cm$^2$, in the low $T$
rhombohedral phase).
The ferroelectric ordering phase 
transition has been characterized as second order and the ordering temperature
decreases, from its value in the ambient pressure sulfide compound,
Se alloying
and also with pressure. \cite{b_studenyak,k_ovsyannikov,zapeka2015critical}
The decrease with Se alloying is potentially significant as it would allow
one to make ferroelectric samples with different ordering temperatures
to characterize for example the effect of proximity to the phase
transition on the PV properties.
Importantly, these are ordered compounds that have been made as
single crystals,
which may facilite characterization and understanding of the interplay of
ferroelectricity and PV properties.

\section{Structures and Methods}

Here we report calculated electronic structures, transport and optical
properties.
Our first principles calculations were performed using the
linearized augmented planewave (LAPW)\cite{LAPW} method as implemented in
WIEN2k code.\cite{wien2k} 
We used LAPW sphere radii of 2.50 Bohr, 1.85 Bohr and 1.95 Bohr for
Sn, P and S in Sn$_2$P$_2$S$_6$
and 2.50 Bohr, 1.86 Bohr and 2.28 Bohr for Sn, P and Se in Sn$_2$P$_2$S$_6$,
respectively.
A basis set cut-off, $k_{max}$,
set by the criterion $R_{min}k_{max}$ = 9.0 was used,
where $R_{min}$ is the P sphere radius.
The experimental lattice parameters are $a$ = 6.529 {\AA}, $b$ = 7.485 {\AA},
$c$ = 11.317 \AA\ and $\beta$ = 124.11$^\circ$ for
$Pc$-Sn$_2$P$_2$S$_6$,\cite{S6stru} and
$a$ = 6.815 {\AA}, $b$ = 7.717 {\AA}, $c$ = 11.694 \AA\ 
and $\beta$ = 124.549$^\circ$ for $Pc$-Sn$_2$P$_2$Se$_6$\cite{Se6stru}.
Note that these are pseudo-orthorhombic structures since
{\bf a} and {\bf a}+{\bf c} are very nearly orthogonal
(angle of $\sim 89^\circ$ in both compounds).

\begin{figure*}
\centerline{\includegraphics[width=1.5\columnwidth]{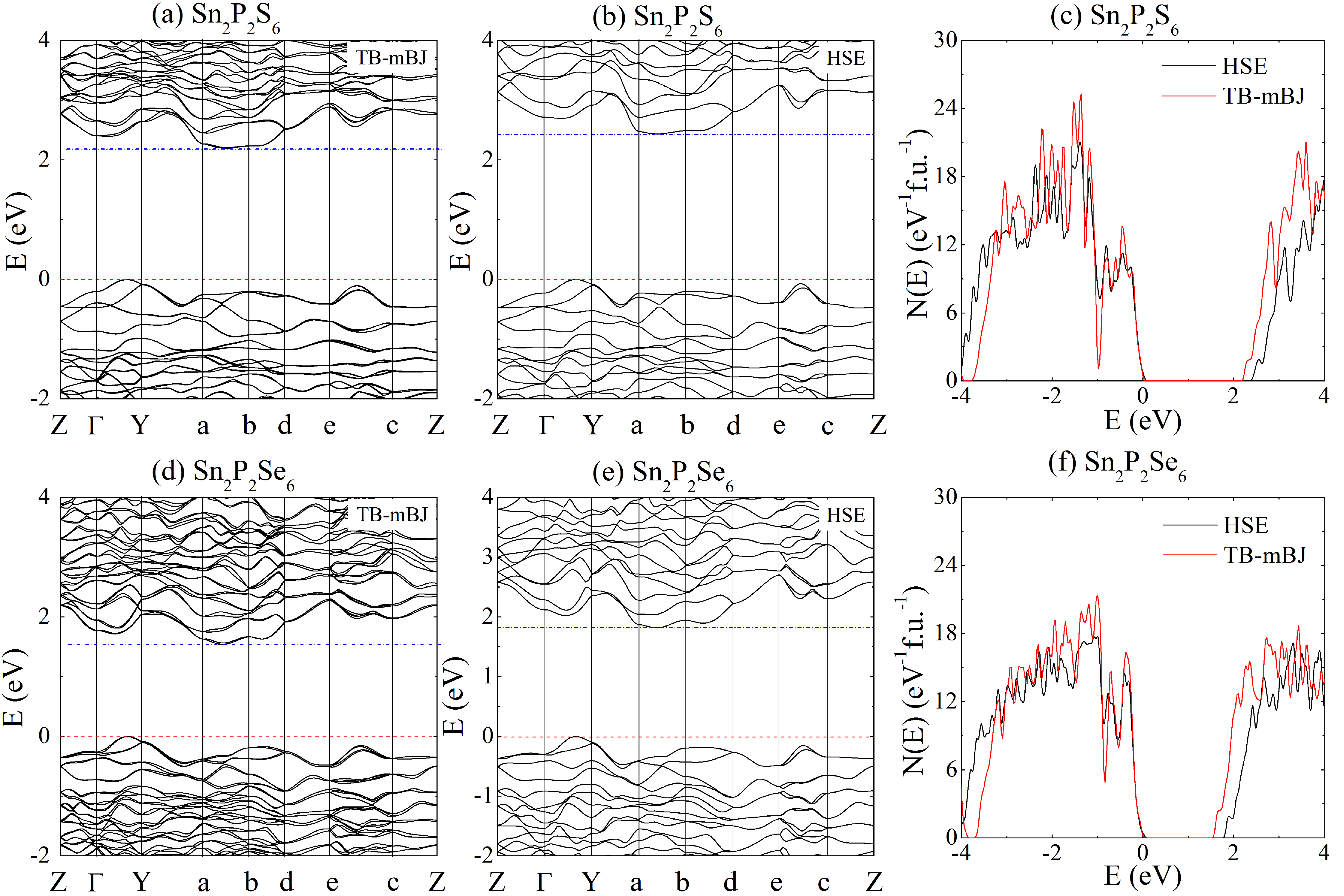}}
\caption{Calculated band structures (a,b,d,e) and densities of states (c,f)
for the two compounds comparing TB-mBJ and HSE results. Note that the
TB-mBJ results include spin-orbit, although as seen
the spin-orbit splittings are small. See text for computational
details. In all cases, energy zero is at the valence band maximum.
}
\label{band}
\end{figure*}

We fixed the lattice parameters to experimental values and relaxed
the internal atomic coordinates using
the Perdew, Burke and Ernzerhof generalized gradient approximation (PBE-GGA).
This resulting structure is polar as in experiment.
Following the structure optimization,
we did electronic structure and optical calculations using the modified
Becke-Johnson type potential of Tran and Blaha,
denoted TB-mBJ in the following.\cite{mBJ1}
This potential gives band gaps in remarkably good accord with experiment for
a wide variety of simple semiconductors and insulators.
\cite{mBJ1,mBJ2,mBJ3,mBJ4,mBJ5}.
However, it should be noted that the TB-mBJ functional has been shown
to disagree with many body calculations for band widths in several
cases. \cite{waroquiers}
We did calculations to cross-check of TB-mBJ results
using the VASP code with the
hybrid HSE06 functional.\cite{HSE01,HSE02}
A comparison of TB-mBJ and HSE
band structures and densities of states is given in Fig. \ref{band}.
As shown, we find very good agreement in the present case.
The most significant difference are in the conduction bands, where
the density of states is slightly broader near the band edge for the HSE
calculation. One may also note a difference in the valence band width,
although this is less significant for the present
study since the shape within $\sim$2 eV of the band edge is very much
the same.

Spin-orbit coupling (SOC) was included in all
TB-mBJ electronic structures and
optical property calculations that follow, although the effects are
small as may be seen from the small spin orbit splittings in the TB-mBJ
band structures (note that the HSE results do not include spin orbit).
Transport coefficients were obtained using the BoltzTrap code.
\cite{madsen2006boltztrap}

\begin{figure}[b!]
\centerline{\includegraphics[width=0.85\columnwidth]{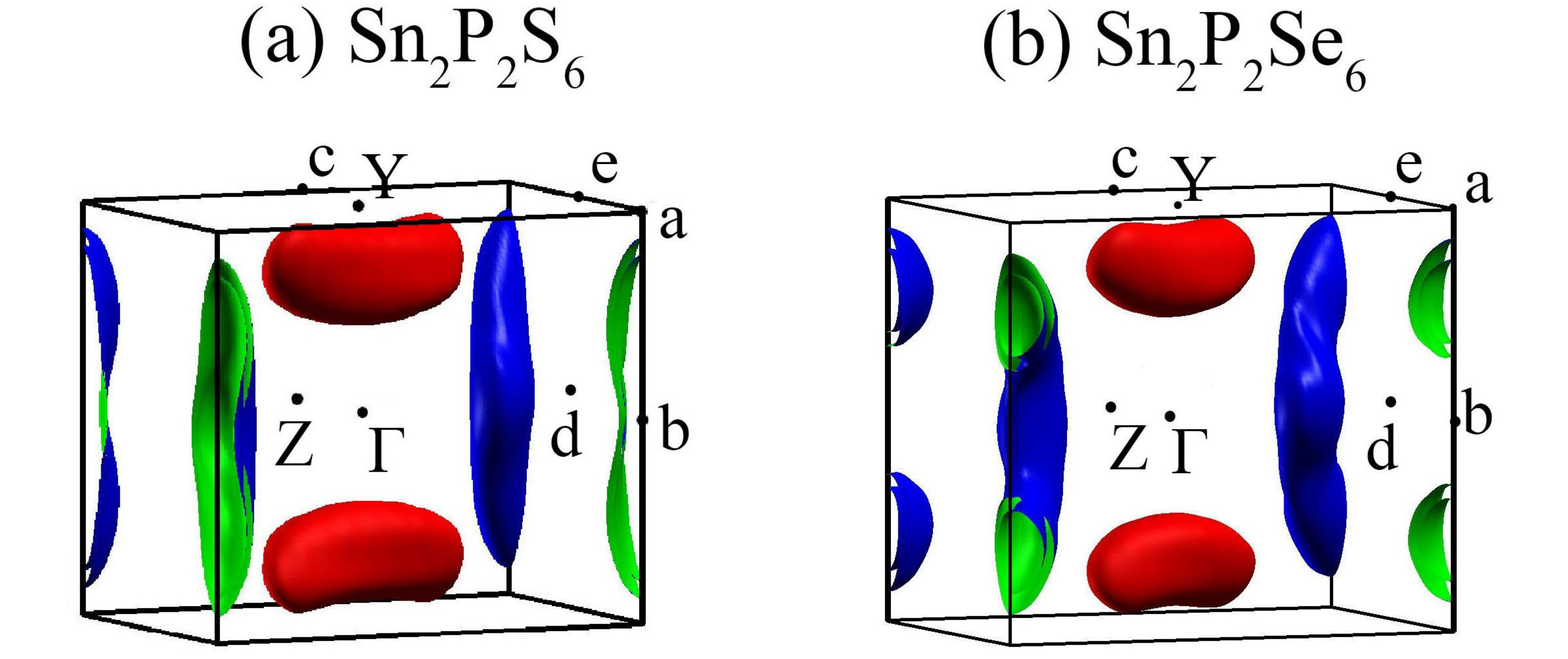}}
\caption{Calculated
isoenergy surfaces illustrating the carrier pockets for Sn$_2$P$_2$S$_6$ (a)
and Sn$_2$P$_2$Se$_6$ (b).
Blue/green is used for the $n$-type isosurface 0.05 eV above CBM
and red is used for the $p$-type isosurface 0.05 eV below VBM.}
\label{fersur}
\end{figure}

\section{Electronic Structures and Optical Absorption} 

\begin{figure}[t!]
\centerline{\includegraphics[width=0.84\columnwidth]{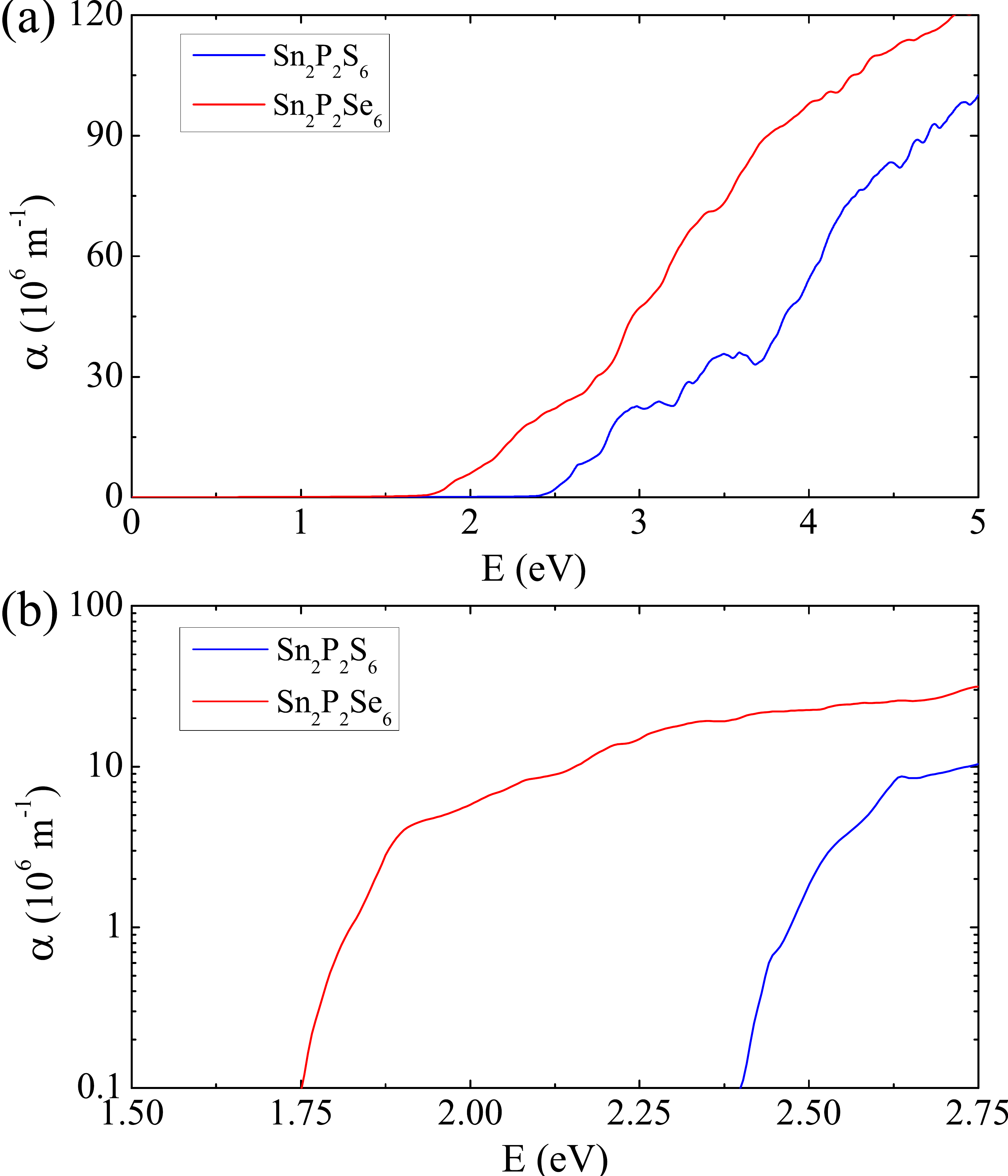}}
\caption{Calculated absorption coefficient for Sn$_2$P$_2$S$_6$ (blue)
and Sn$_2$P$_2$Se$_6$ (red) with normal scale between 0 and 5 eV (a)
and with log scale between 1.5 and 2.75 eV (b).
We use very low broadening to make the features more apparent.
These are 0.02 eV and 0.005 eV, for (a) and (b), respectively.
Note in particular the features near 2.6 eV, 3 eV and 3.5 eV for the
sulfide}
\label{absorb}
\end{figure}

We begin with the electronic structures of
Sn$_2$P$_2$S$_6$ and Sn$_2$P$_2$Se$_6$. Fig. \ref{band} shows the band
structures, and the Brillouin zones with high symmetry points and the
carrier pockets are shown in Fig. \ref{fersur}.
Both compounds are indirect gap materials.
The band gaps of Sn$_2$P$_2$S$_6$ and Sn$_2$P$_2$Se$_6$ are 2.20 eV and 1.55 eV respectively.
The valence band maxima (VBM) are between $\Gamma$ and Y points and
the conduction band minima (CBM) are between a and b points.
The direct band gaps of Sn$_2$P$_2$S$_6$ and Sn$_2$P$_2$Se$_6$ are
2.43 eV and 1.76 eV respectively.
We find strong optical absorption in Sn$_2$P$_2$S$_6$
for short wavelength visible light
($\sim \lambda$ \textless\ 560 nm) and in Sn$_2$P$_2$Se$_6$
for the visible range, as seen in the Fig. \ref{absorb} (a) and (b).
This starts at the absorption edges of 
2.40 eV and 1.75 eV, for the sulfide and selenide, respectively.

It is to be noted that in general direct gap semiconductors are preferred
in PV applications as the voltage is limited by the fundamental indirect
gap. However, photons with energies above the indirect gap but below the
direct gap are typically lost due to insufficient
absorption leading to reduced efficiency
(Si PV cells are an exception). In the present context, the difference between
the direct and indirect gaps is not large, and therefore in an actual
PV application only a small loss
in efficiency due to the indirect gap would be expected, and this might be
compensated by reduced electron-hole recombination, also as a result of the
indirect nature of the gap.

As mentioned, there have been a number of experimental
studies of the absorption
at the band edges of these compounds.
We compare the experimental band gaps at low temperature with our calculated
results in Table \ref{gap}. Reasonable agreement is found.
Turning to the region above the band edge, Gamernyk and co-workers
\cite{j_gamernyk}
reported photoconductivity and photoluminescence spectra for Sn$_2$P$_2$S$_6$
crystals up to 3.5 eV. Both spectra show structures at $\sim$ 2.7 eV,
$\sim$3.0 eV and $\sim$3.4 eV. Our absorption spectrum for
Sn$_2$P$_2$S$_6$ also shows three main features between the onset
ant 4 eV. These are a peak at 2.6 eV, a broader peak (with substructure)
at $\sim$ 3 eV and another broad peak, again with substructure at $\sim$
3.5 eV, in close
correspondence with the features in the experimental spectra.

\begin{table}[b]
\centering
\caption{Direct/indirect gaps compared with low-temperature experimental data.
}
\resizebox{8.6cm}{!}{
\begin{tabular}{c|c|c||c|c} 
\hline\hline
\multirow{2}{*}{} & \multicolumn{2}{c||}{Sn$_2$P$_2$S$_6$}                & \multicolumn{2}{c}{Sn$_2$P$_2$Se$_6$}    \\
\cline{2-5}   
     & Direct gap         & Indirect gap   &Direct gap       & Indirect gap  \\
     &  (eV)              &      (eV)      &  (eV)           &      (eV)     \\
\hline 
Our work       &   2.43      & 2.20       &   1.76      & 1.55              \\
\hline
A. Ruediger\cite{o_ruediger}&   2.50  &  -    &   -      &   -          \\
\hline
Z. Pot{\.{u}}{\v{c}}ek\cite{ao_potuuvcek2004luminescence} &2.50 &  -  &    -                                            &   -                      \\
\hline
E. Gerzanich\cite{r_gerzanich}  &    $ \approx $2.65     &  -                        &   $ \approx $2.00                       &   -                      \\
\hline    
J. Lipavi{\v{c}}ius\cite{q_lipavivcius}       &     -         &       -                   &   $ \approx $2.10                      &  -                        \\
\hline 
\end{tabular}
}
\label{gap}
\end{table}

\begin{figure}[t!]
\centerline{\includegraphics[width=0.85\columnwidth]{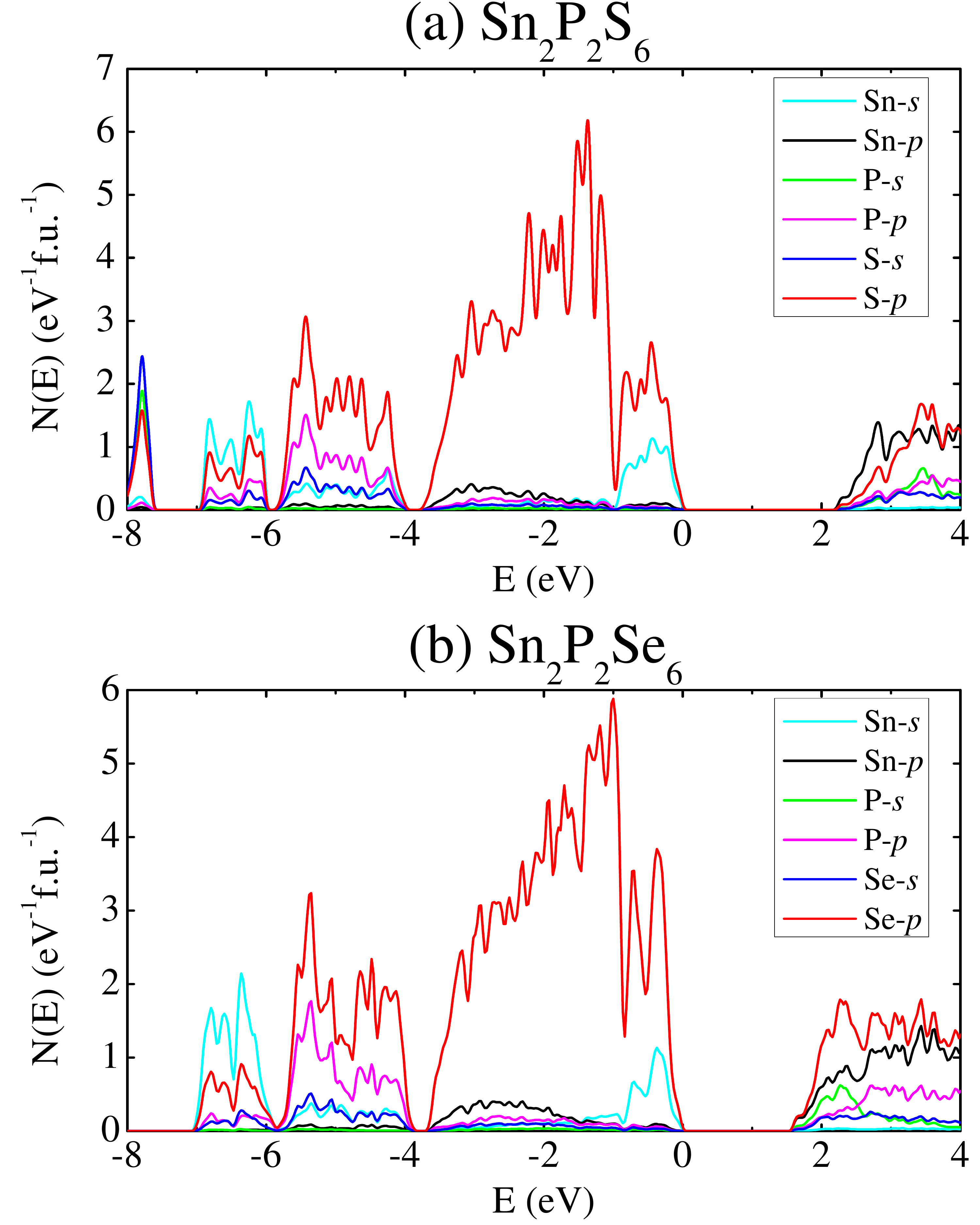}}
\caption{Projected density of states of Sn$_2$P$_2$S$_6$ (a) and Sn$_2$P$_2$Se$_6$ (b). The VBM is at zero.
}
\label{dos}
\end{figure}

The projected densities of states (pDOS), shown in Fig. \ref{dos},
indicate the VBM of Sn$_2$P$_2$S$_6$ and Sn$_2$P$_2$Se$_6$ are mainly dominated by S-3$p$ and Se-4$p$ orbitals respectively.
Especially,
there are separate energy ranges for Sn-5$s$/S-3$p$ and Sn-5$s$/Se-4$p$
coupling between -7 eV and -6 eV and between -1 eV and 0 eV.
This shows divalent Sn as expected.\cite{SnO, li2016design}
It also shows Sn-5$s$/S-3$p$ (and  Sn-5$s$/Se-4$p$) bonding states between -7 eV
and -6 eV and Sn-5$s$/S-3$p$ (and  Sn-5$s$/Se-4$p$) antibonding states
between -1 eV and 0 eV.
Anti-bonding character at a VBM usually causes defect tolerant behavior,
\cite{zhang1998defect, brandt2015identifying}
i.e., bond breaking associated with the formation of defect states
will produce shallow rather than deep acceptor levels in the mid-gap region.
This greatly facilitates $p$-type doping,
giving rise to ambipolar conductivity in photovoltaic materials,
such as chalcopyrites\cite{herberholz1997prospects}.
Previous research indicates the feasibly of
$p$-type conductivity for Sn$_2$P$_2$S$_6$\cite{golden2016sn}
and considering the present results it also seems likely for
Sn$_2$P$_2$Se$_6$ with suitable dopants or metal vacancies.

\begin{figure}[t!]
\centerline{\includegraphics[width=0.85\columnwidth]{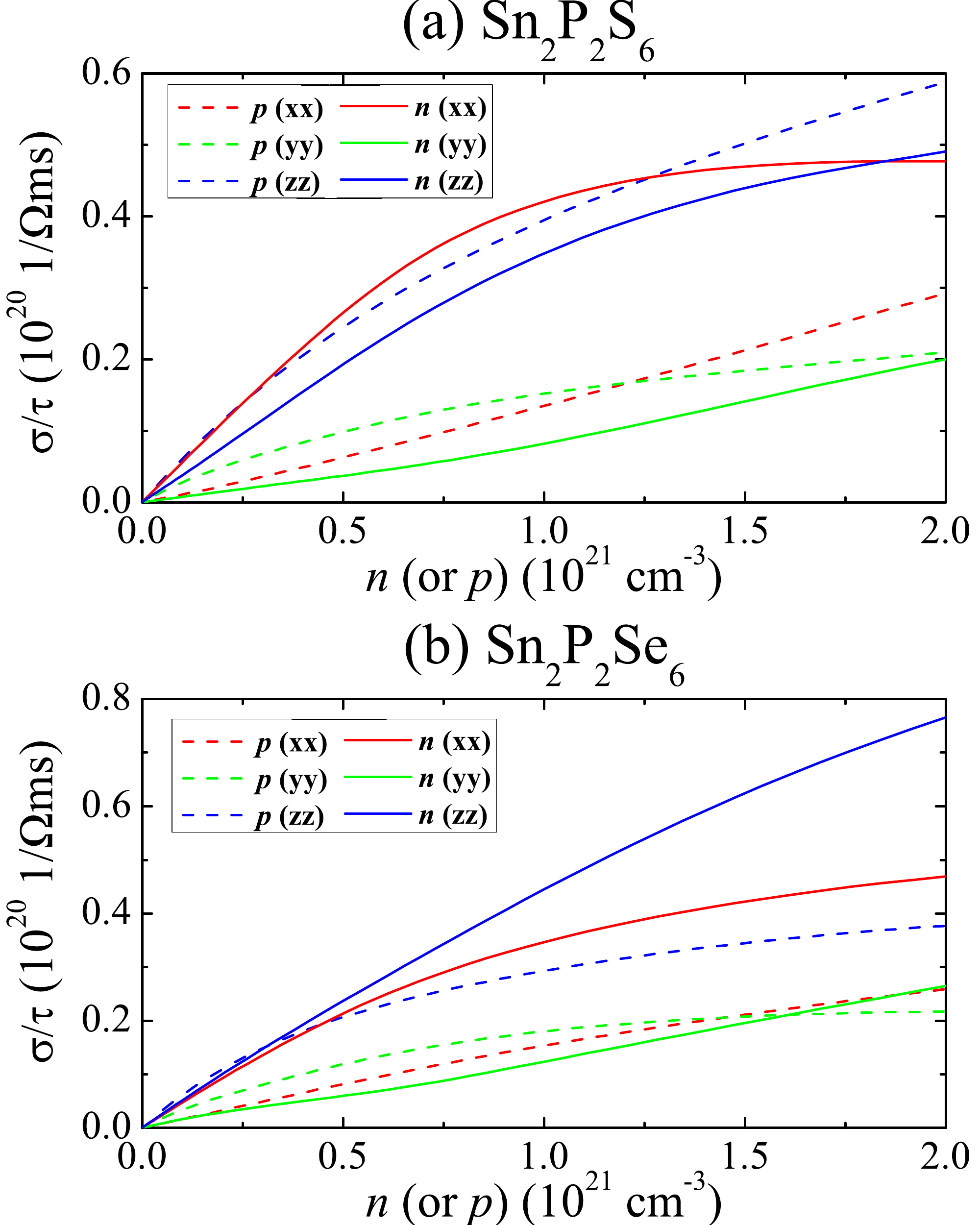}}
\caption{Calculated anisotropic $\sigma$/$\tau$ of Sn$_2$P$_2$Se$_6$ (a)
and Sn$_2$P$_2$Se$_6$ (b)
as a function of carrier concentration $n$ (or $p$) at 300 K in xx (red),
yy (green) and zz (blue) directions.
Solid lines mean $n$-type and dash lines mean $p$-type.
}
\label{cond}
\end{figure}

\section{Transport Related Properties}

\begin{table}[b!]
\centering
\caption{Direction average and and components of the electron (m$^* _e$)
and hole (m$^* _h$) transport effective mass, obtained from the
calculated $\sigma/\tau$ at carrier concentration 10$^{18}$ cm$^{-3}$.}
\resizebox{5.5cm}{!}{
\begin{tabular}{c|c|c|c|c|c} 
\hline\hline
\multicolumn{2}{c|}{}                                                & average                   &  xx            &                   yy                    &      zz     \\
\hline
\multirow{2}{*}{Sn$_2$P$_2$S$_6$} & m$^* _e$ &  0.82                       & 0.50           &                   3.61                 &     0.74   \\
\cline{2-6}
                                                                &  m$^* _h$ &  0.81                       &2.86           &                   0.96                 &      0.43   \\
\hline 
\multirow{2}{*}{Sn$_2$P$_2$Se$_6$} & m$^* _e$&  0.71                       & 0.59         &                   1.53                 &     0.54   \\
\cline{2-6}
                                                                &  m$^* _h$ &  0.69                       &1.83           &                   0.73                 &      0.41   \\
\hline
\end{tabular}
}
\label{mass}
\end{table}

The calculated anisotropic transport function, $\sigma$/$\tau$, of Sn$_2$P$_2$S$_6$ and Sn$_2$P$_2$Se$_6$ at 300 K is shown in Fig. \ref{cond}.
Significantly anisotropic $n$-type and $p$-type transport is found.
For $n$-type Sn$_2$P$_2$S$_6$,
$\sigma$/$\tau$ in x and z directions are similar and both are larger than
that in y direction (the Cartesian directions are x along {\bf a},
y along {\bf b} and z normal to the x and y directions, very close to
the pseudoorthorhombic lattice direction {\bf a}+{\bf c}).
The $p$-type $\sigma$/$\tau$ of Sn$_2$P$_2$S$_6$ in z direction is large,
even better than $n$-type $\sigma$/$\tau$ for some carrier concentrations.
For Sn$_2$P$_2$Se$_6$, the $n$-type and $p$-type $\sigma$/$\tau$
in all directions are similar, except the $n$-type $\sigma$/$\tau$ in z
direction which is larger than that in other directions.
So the electron $\sigma$/$\tau$ is larger than the hole $\sigma$/$\tau$ for
Sn$_2$P$_2$S$_6$.
The $c$-axis direction has the most favorable transport with high
$\sigma$/$\tau$ for both electrons and holes.
The effective masses of them have the same anisotropic characteristics
with transport properties as seen in the Table \ref{mass}.
The relatively low m$^* _e$ and m$^* _h$ suggest reasonable carrier
transport.

$n$/$p$-type Sn$_2$P$_2$S$_6$ and Sn$_2$P$_2$Se$_6$ have remarkably
good $\sigma$/$\tau$ for materials with complex crystal structures. As can be seen from the parabolic band expression $\sigma$/$\tau$ $\propto$  $n/m^*$, the origin of the good $\sigma$/$\tau$ for Sn$_2$P$_2$S$_6$ and Sn$_2$P$_2$Se$_6$ can be understood in terms of the detailed band structure. 
For $p$-type Sn$_2$P$_2$S$_6$ and Sn$_2$P$_2$Se$_6$, the VBM is derived
from two bands, which come from the antibonding coupling of S (or Se)
$p$ orbitals with Sn $s$ orbitals.
The antibonding coupling increases the upper valence band dispersion,
as seen in the Fig. \ref{band},
decreases the density of states, as seen in  the Fig. \ref{dos},
and leads to complex shaped isoenergy surfaces at relatively low carrier
concentrations as depicted in Fig. \ref{fersur}.
For $n$-type
the CBM also comes from dispersive bands and shows complex shaped isosurfaces
even at relatively low carrier concentrations that consist of
elongated sections (pipes) running along zone edges.
These are very far from spherical carrier pockets.
This type of isosurface structure,
consisting of cylinders or curved cylinders,
is also present in cubic $p$-type PbTe\cite{singh2010doping}
and $n$-type SrTiO$_3$\cite{sun2016thermoelectric},
which are also materials that show remarkably good conductivity
and are near ferroelectricity.
As discussed for those materials,
this leads to the combination of a relatively light transport effective
mass (for conductivity) and heavy density of states effective mass.
In the present case light mass dominated by the dispersion transverse
to the cylinders enters the conductivity and a heavier mass given by an
average enters the density of states.
\cite{parker2013high,shirai2013mechanism,bilc,sun2016thermoelectric,shi,xing}
The bottom conduction band is highly dispersive across the cylinder directions, contributing strongly to the conductivity,
and weakly dispersive along the cylinders,
contributing to a high density of states, and thus joint density of states
favoring strong absorption.

\section{Discussion and Conclusions}

In summary, we report the electronic structures, transport
and optical properties of ferroelectric Sn$_2$P$_2$S$_6$ and Sn$_2$P$_2$Se$_6$
based on first principles calculations.
They have large visible light absorption and band structures, as
seen in the carrier pocket shapes and transport effective masses, consistent
with relatively good transport properties.
Therefore these compounds are useful model systems for investigating the
interplay of ferroelectricity, photoresponse and charge collection.
Besides use for fundamental investigation of PV effects in ferroelectrics,
it would also be of interest to investigate them as potential practical
solar absorbers. This is because of the reasonable band gap,
especially for the selenide.
While the selenide is not ferroelectric at room temperature,
the high dielectric constant associated with nearness to
ferroelectricity and the favorable electronic structure may lead to
high defect tolerance and favorable carrier collection.
In this regard we note that alloys of the sulfide and selenide have
been reported in which Curie temperature can be tuned to room temperature.
\cite{zapeka2015critical} 
Investigation of this alloy may then be useful for
studying changes in charge collection as one passes
through a ferroelectric transition near room temperature.
In any case, the present results suggest investigation of crystalline
Sn$_2$P$_2$S$_6$ and Sn$_2$P$_2$Se$_6$ in order to develop understanding
of ferroelectric PV effects in semiconductors.

\acknowledgements

This work was partly supported by the Department of Energy through
the MAGICS center, award DE-SC0014607 (YL, first principles).
DJS acknowledges support from the Department of Energy, S3TEC EFRC,
award DE-SC0001299/DE-FG02-09ER46577 (optical properties).

\bibliography{reference}

\end{document}